\documentclass{iaus}
\usepackage{graphicx}
\usepackage{url}
\usepackage{natbib}

\title[Evolution of sunspot magnetic fields] %% give here short title %%
{Automated sunspot detection and the evolution of sunspot magnetic fields\\during solar cycle 23}

\author[Fraser Watson \& Lyndsay Fletcher]   %% give here short author list %%
{Fraser Watson
 \and Lyndsay Fletcher}

\affiliation{Department of Physics and Astronomy \\ Kelvin Building,
University of Glasgow, Glasgow, UK \\ email: {\tt f.watson@astro.gla.ac.uk}}

\pubyear{2010}
\volume{273}  %% insert here IAU Symposium No.
\pagerange{119--126}
\jname{Physics of Sun and Star Spots}
\editors{A.C. Editor, B.D. Editor \& C.E. Editor, eds.}
\begin{document}

\maketitle

\begin{abstract}
The automated detection of solar features is a technique which is relatively underused but
if we are to keep up with the flow of data from spacecraft such as the recently launched
Solar Dynamics Observatory, then such techniques will be very valuable to the solar
community. Automated detection techniques allow us to examine a large set of data in
a consistent way and in relatively short periods of time allowing for improved statistics 
to be carried out on any results obtained. This is particularly useful in the field of 
sunspot study as catalogues can be built with sunspots detected and tracked without any 
human intervention and this provides us with a detailed account of how various sunspot properties
evolve over time. This article details the use of the Sunspot Tracking And Recognition Algorithm 
(STARA) to create a sunspot catalogue. This catalogue is then used to analyse the magnetic fields
in sunspot umbrae from 1996-2010, taking in the whole of solar cycle 23.

\keywords{Sun: evolution - Sun: magnetic fields - Sun: photosphere - sunspots - techniques: image processing}
\end{abstract}

\firstsection % if your document starts with a section,

\section{Introduction}
To examine the magnetic fields measured in sunspots it is useful to have large datasets as there are vast differences between a simple sunspot surrounded by quiet sun and a sunspot which is in the centre of a complex active region. The large dataset was assembled by using an automated sunspot detection algorithm developed by \citet{Watson2009}. The Sunspot Tracking And Recognition Algorithm (STARA) is a quick and reliable way to process a large number of solar images and has been tested on images from a variety of sources including ground based observatories (such as Kanzelh\"{o}he Observatory, see \url{http://www.kso.ac.at/sonnenbeobachtung/spot_rec_en.php} for details on how the algorithm is being used), the MDI instrument on SOHO and the HMI instrument on the SDO satellite.

The data used in this article were level 1.8 data recorded by the MDI instrument \citep{Scherrer1995} and are taken from the launch of the instrument in 1996 through to early 2010 which covers the whole of solar cycle 23. We use both white light continuum observations and magnetograms which allows us to detect the sunspots and determine their magnetic properties at the same time.

\section{Creating the Catalogue}
To ensure that the magnetic fields of sunspots could be measured at the same time as their detection, the times of measurement had to be as close as possible. Due to the cadence of the MDI measurements, this was only the case once per day at 00:00UT giving a dataset of around 5000 continuum and magnetogram images to process. Processing these images takes 30-40 hours on a single processor depending on the number of sunspots present. To extract the sunspots from the data, techniques from the field of mathematical morphology were used (see \citet{Matheron1975} and \citet{Serra1982} for more detail).

%\begin{figure}[htb]
%\begin{center}
%\includegraphics[width = 0.4\textwidth]{transformexplain1.eps}
%\caption{An example solar continuum image from MDI. The black horizontal line passes through 2 sunspots.}
%\label{fig:sun}
%\end{center}
%\end{figure}

\begin{figure}[htb]
	\parbox[b]{.4\linewidth}{
	\begin{center}
		\includegraphics[width = 0.4\textwidth]{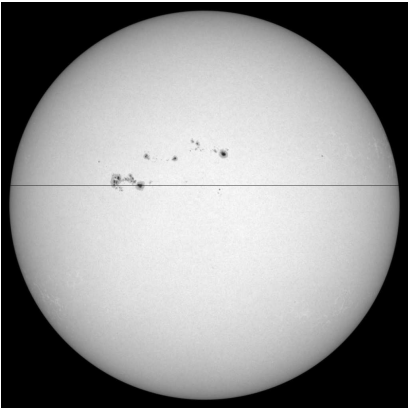}\end{center}}\hfill
	\parbox[b]{.56\linewidth}{
		\caption{An example solar continuum image from MDI. The black horizontal line passes through 2 sunspots.}
		\label{fig:sun}
	}
\end{figure}

%\begin{figure}[htb]
%\begin{center}
%\includegraphics[width = 0.8\textwidth]{transformexplainarraybw2.eps}
%\caption{The various image processing steps involved in sunspot detection with the STARA algorithm. Top left: inverted profile from the horizontal line in Figure~\ref{fig:sun} (the two peaks show the two sunspots on that line), top right: after erosion, bottom left: after dilation, bottom right: after subtraction of original profile.}
%\label{fig:explain}
%\end{center}
%\end{figure}

\begin{figure}[htb]
	\parbox[b]{.2\linewidth}{
	\begin{center}
		\caption{The various image processing steps involved in sunspot detection with the STARA algorithm. Top left: inverted profile from the horizontal line in Figure~\ref{fig:sun} (the two peaks show the two sunspots on that line), top right: after erosion, bottom left: after dilation, bottom right: after subtraction of original profile.}\end{center}}\hfill
	\parbox[b]{.78\linewidth}{
		\includegraphics[width = \linewidth]{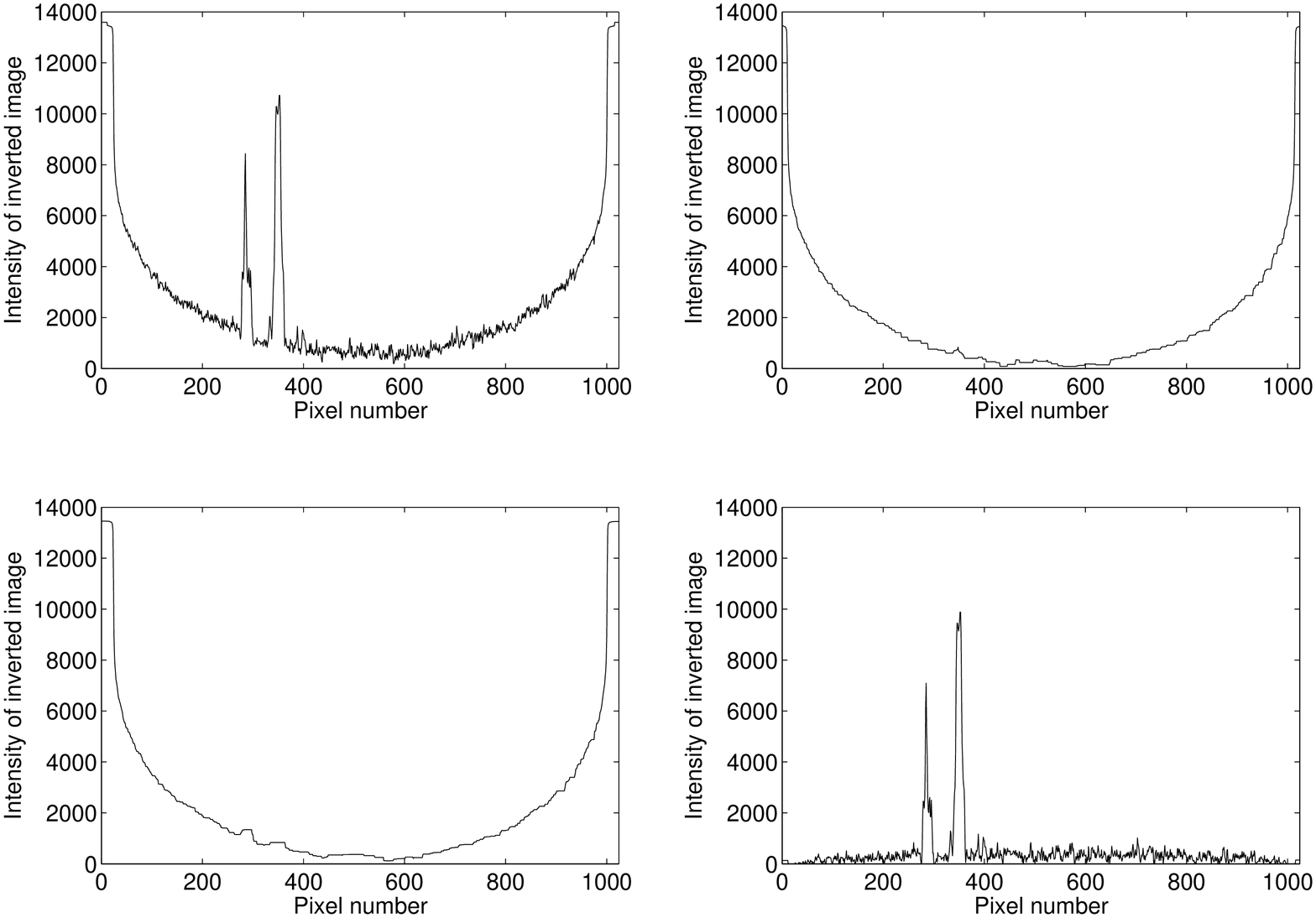}

		\label{fig:explain}
	}
\end{figure}

Figure~\ref{fig:sun} and Figure~\ref{fig:explain} shows the various steps involved in detecting sunspots using the STARA code and the process works as follows (note that this example is given in 2D for simplicity looking at the two sunspots along the dark line in Figure~\ref{fig:sun} but the same applies to the full 3D Sun case) :

\begin{itemize}
\item Invert the image so that the sunspots appear as bright peaks on a darker background. This is shown in the top left panel of Figure~\ref{fig:explain}.
\item A `top-hat' transform is then applied which consists of an erosion and a dilation (explanations of these terms can be found in Serra, 1982).
\item Subtract the profile after this transform from the original to obtain the bottom right panel of Figure~\ref{fig:explain}.
\item Apply a threshold to give locations of the sunspots.
\end{itemize}

The full 3D case works in the same way but rather than operating on a 2D `U' shaped profile, a 3D bowl shaped profile is used. Further processing is then applied to separate the umbra and penumbra of the sunspots. To complete this process on a single image takes around 4 seconds. The sunspot locations are then superimposed on a magnetogram taken at almost the same time so that the magnetic fields can be recorded. This is repeated until all of the images have been processed.

\section{Magnetic fields in sunspots from 1996-2010}

As the magnetic fields present in sunspots had already been measured, we could easily look at the trends present over the length of the catalogue.

%\begin{figure}[b]
%\begin{center}
%\includegraphics[width = 0.7\textwidth]{raw_spot_data.eps}
%\caption{Maximum sunspot umbra fields from 1996 to 2010 as measured by the STARA algorithm.}
%\label{fig:Watsonfields}
%\end{center}
%\end{figure}

\begin{figure}[b]
	\parbox[b]{.7\linewidth}{
		\includegraphics[width = \linewidth]{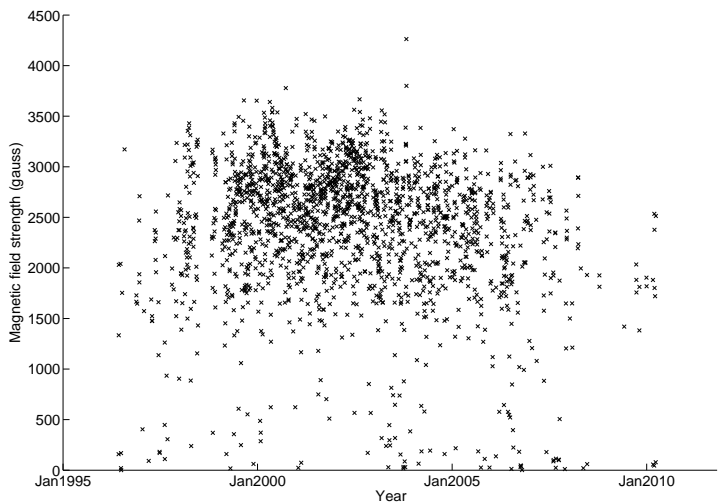}}\hfill
	\parbox[b]{.26\linewidth}{
	\begin{center}
		\caption{Maximum sunspot umbra fields from 1996 to 2010 as measured by the STARA algorithm.}
		\end{center}
		\label{fig:Watsonfields}
	}
\end{figure}

Figure~\ref{fig:Watsonfields} shows the maximum magnetic field detected in the umbra of all sunspots present on a given day. It is assumed that the field in the sunspot umbrae are in the local vertical direction and so a cosine correction is applied to the MDI line of sight magnetic field. To minimise the effects of this, only sunspots with a value of $\mu > 0.95$ were used (where $\mu$ is the cosine of the angle between the local solar vertical and the observers line of sight). Also, if there is a day with no sunspots, that day is omitted from the plot.

We can immediately see that there is a large variation in the sunspot fields even over short timescales of a few weeks and that the majority of maximum umbral fields fall between 1500 and 3500 gauss.

\begin{figure}[t]
	\parbox[b]{.65\linewidth}{
		\includegraphics[width = \linewidth]{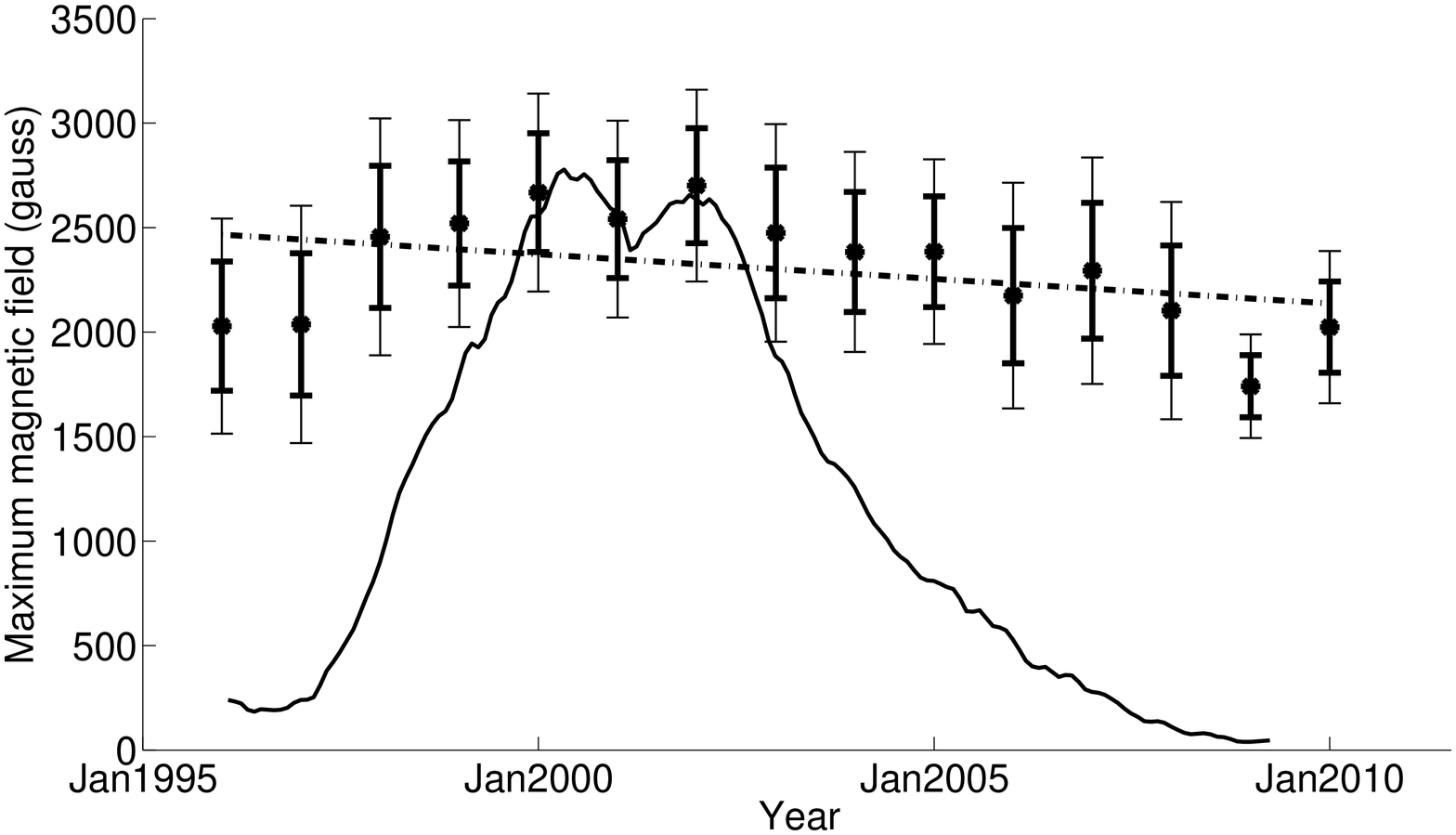}}\hfill
	\parbox[b]{.31\linewidth}{
		\caption{The data from Figure~\ref{fig:Watsonfields} have been binned by year and the mean of each bin plotted. The dashed line is the best fit straight line to those points. The gradient corresponds to a trend of -23.6 gauss per year. Thin error bars correspond to the standard deviation of all data in Figure~\ref{fig:Watsonfields} whereas thick error bars only take into account the data which met the criteria for being a sunspot in the \citet{Penn2006} article. The solid line is the scaled international sunspot number over the same time period shown for reference.}
		\label{fig:means}
	}
\end{figure}

\citet{Penn2006} looked at this long term trend using the McMath-Pierce telescope on Kitt Peak and measured the magnetic field in the darkest observed part of sunspots from around 1993 to the present. This was done by measuring the Zeeman splitting present in the Fe I line to infer a magnetic field strength at the location of the measurement. When their whole data set is taken into account they find a trend of decreasing sunspot magnetic field strength of about 52 Gauss per year which is obtained by binning the data by year and looking at the mean of each bin along with the standard error on the mean. In Figure~\ref{fig:means} we show the same treatment of the data from the STARA algorithm. The best fitting straight line to the data is shown with a dashed line and there are two sets of error bars present. The thin error bars show the standard deviation of all the data in that bin and the thick error bar only takes into account data which would meet the criteria for being a sunspot by \citet{Penn2006}. This is because their data excludes pores, some of which can be as large as 10" in size and we observe sunspots that are of that order in size. These sunspots are the primary reason for the fields in Figure~\ref{fig:Watsonfields} that lie below 1500 gauss as they correspond to small sunspots that do not yet have magnetic field strengths comparable to larger, fully formed spots.

%\begin{figure}[t]
%\begin{center}
%\includegraphics[width = 0.65\textwidth]{mean_stdev_fit.eps}
%\caption{The data from Figure~\ref{fig:Watsonfields} have been binned by year and the mean of each bin plotted. The dashed line is the best fit straight line to those points. The gradient corresponds to a trend of -23.6 gauss per year. Thin error bars correspond to the standard deviation of all data in Figure~\ref{fig:Watsonfields} whereas thick error bars only take into account the data which met the criteria for being a sunspot in the \citet{Penn2006} article. The solid line is the scaled international sunspot number over the same time period shown for reference.}
%\label{fig:means}
%\end{center}
%\end{figure}

We observe the same decreasing trend as \citet{Penn2006} but with a shallower gradient of 23.6 gauss per year. However, it could be argued that there is a slight cyclic variation - unfortunately we have no cycle 22 data with which to investigate this. The data from the STARA algorithm also has a much larger spread making the errors larger in this result. However, the advantage of making measurements in this way is that it allows for a completely automated system that always processes data in the same way. From this data we cannot say for certain that a long term trend exists until data from the new solar cycle is obtained and a scaled plot of the international sunspot number is included in Figure~\ref{fig:means} to show that the mean magnetic field is increasing and decreasing along with the solar activity. A change of 600G over the solar cycle, as suggested by \citet{Penn2006} would cause a change in the mean umbral radius as a relationship has been shown by \citet{Kopp1992} and \citet{Schad2010} but observations by \citet{Penn2007} could not uncover this in the data. \citet{Mathew2007} also suggests that the size distribution of sunspots, although constant over the solar cycle, could introduce a bias into small sunspot samples if the size distribution of spots is not calculated.

\begin{acknowledgements}
FW acknowledges the support of an STFC PhD studentship. LF acknowledges 
the support of the EC-funded SOLAIRE Research and Training Network 
(MTRN-CT-2006-035484), the STFC  (Rolling Grant ST/F002637/1) and the 
Leverhulme Foundation (Grant F/00 179/AY). We also acknowledge Prof. Stephen Marshall of the University of Strathclyde for useful information and discussion regarding the image processing techniques used.
\end{acknowledgements}

\end{document}